\documentclass[aps,prl,amsmath,amssymb,amsfonts,superscriptaddress,floatfix]{revtex4}
\usepackage{amsmath}
\usepackage{graphicx}
\usepackage{amsthm}
\newtheorem{theorem}{Theorem}
\newtheorem{lemma}{Lemma}
\newcommand{\cmm}{c_{MM}}
\newcommand{\ME}{\Psi_{\rm ME}}
\newcommand{\cmax}{\chi_{\rm max}}
\newcommand{\dSm}{\delta S^{\rm max}}

\newcommand{\conc}{{\cal E}^C}
\newcommand{\Pb}{P_{bad}}
\newcommand{\pN}{P_{near}}

\newcommand{\be}{\begin{equation}}
\newcommand{\ee}{\end{equation}}
\usepackage{multicols}

\newcommand{\Pc}{P_{same}}
\newcommand{\FF}{F}

\begin{document}
\title{Superadditivity of communication capacity using entangled
inputs}
\author{M.~B.~Hastings}
\affiliation{Center for Nonlinear Studies and Theoretical Division,
Los Alamos National Laboratory, Los Alamos, NM, 87545}
\maketitle

\begin{multicols}{2}
{\bf
The design of error-correcting codes used in modern communications relies
on information theory to quantify the capacity of a noisy channel to send
information\cite{infthy}.  This capacity can be expressed using the mutual information
between input and output for a single use of the channel:
although correlations between
subsequent input bits are used to correct errors, they cannot increase the
capacity.  For quantum channels, it has been an open question whether
entangled input states can increase the capacity to send classical
information\cite{holevo}.  The additivity conjecture\cite{moec,equiv} states that entanglement
does not help, making practical computations of the capacity possible.
While additivity is widely believed to be true, there is no proof.  Here
we show that additivity is false, by constructing a random counter-example.
Our results show that the most basic question of classical capacity of a
quantum channel remains open, with further work needed to determine in
which other situations entanglement can boost capacity.}

In the classical setting, Shannon presented a formal definition of a noisy channel ${\cal E}$
as a probabilistic map from input states to output states.  In the quantum
setting, the channel becomes a linear, completely positive, trace-preserving
map from density matrices to
density matrices, modeling noise in the system due to interaction
with an environment.
Such a channel can be used to send either quantum or classical information.
In the first case, a dramatic violation
of {\it operational additivity} was recently shown, in that
there exist two channels, each of which has zero capacity to send
quantum information no matter how many times it is used, but which can be used in tandem to send quantum information\cite{jy}.

Here we address the classical capacity of a quantum channel.
To specify how information is encoded in the channel,
we must pick a set of states $\rho_i$ which we use as input
signals with
with probabilities
$p_i$.
Then the Holevo formula\cite{holevo} for the capacity is:
\be
\label{Hm}
\chi=H\Bigl(\sum_i p_i {\cal E}(\rho_i)\Bigr)-\sum_i p_i H\Bigl({\cal E}(\rho_i)\Bigr),
\ee
where $H(\rho)=-{\rm Tr}(\rho \ln(\rho))$ is the von Neumann entropy.
The maximum capacity of a channel is the maximum over all
input ensembles:
\be
\label{chimax}
\cmax({\cal E})={\rm max}_{\{p_i\},\{\rho_i\}}\chi({\cal E},\{p_i\},\{\rho_i\}).
\ee
Suppose we have two different channels, ${\cal E}_1,{\cal E}_2$.
To compute this capacity, it seems necessary
to consider entangled input states between the two channels.   Similarly,
when using the same channel multiple times, it may be useful to
use input states which are entangled across multiple
uses of the same channel.
The additivity conjecture (see Figure 1) is the conjecture that this does
not help and that instead
\be
\cmax({\cal E}_1 \otimes {\cal E}_2) = \cmax({\cal E}_1)+
\cmax({\cal E}_2).
\ee

The additivity conjecture makes it possible
to compute the classical capacity of a quantum channel.
Further, Shor\cite{equiv} showed that several different additivity conjectures
in quantum information theory
are all equivalent.  These are the additivity conjecture for the
Holevo capacity, the additivity conjecture for entanglement of formation\cite{eof}, strong
superadditivity of entanglement of formation\cite{sa},
and the additivity conjecture for minimum output entropy\cite{moec}.  In this
Letter, we show that all of these conjectures are false, by
constructing a counterexample to the last of these conjectures.
Given a channel ${\cal E}$, define the minimum output entropy $H^{\rm min}$ by
\be
H^{\rm min}({\cal E})={\rm min}_{|\psi\rangle} H({\cal E}(|\psi\rangle\langle\psi|)).
\ee
The minimum output entropy conjecture is that for all channels ${\cal E}_1$ and ${\cal E}_2$, we
have
\be
H^{\rm min}({\cal E}_1 \otimes {\cal E}_2)=H^{\rm min}({\cal E}_1)+H^{\rm min}({\cal E}_2).
\ee
A counterexample to this conjecture would be an
entangled input state which has a lower output entropy, and hence
is {\it more} resistant to noise, than any
unentangled state (see Figure 2).

Our counterexample to the additivity of minimum output
entropy is based on a random construction,
similar to those Winter and Hayden used to
show violation of the maximal $p$-norm multiplicativity conjecture for all
$p>1$\cite{aw,ph,phaw}.  For $p=1$,
this violation would imply violation of the minimum output entropy conjecture; however, the
counterexample found in \cite{ph} requires a matrix size which diverges as $p\rightarrow 1$.
We use different system and environment sizes (note that $D<<N$ in our construction
below) and make a different analysis of the probability of different output entropies.
Other violations are known 
for $p$ close to $0$\cite{smallp}.

We define a pair of channels ${\cal E}$ and $\overline {\cal E}$ which
are complex conjugates of each other.  Each channel
acts by randomly choosing a unitary from a small set of unitaries $U_i$
($i=1...D$) and applying that to $\rho$.  This models a situation
in which the unitary evolution of the system
is determined by an unknown state of the environment.
We define
\begin{eqnarray}
{\cal E}(\rho)&=&
\sum_{i=1}^D 
P_i U^{\dagger}_i \rho U_i,
\\ \nonumber
\overline {\cal E}(\rho)&=&
\sum_{i=1}^D 
P_i \overline U^{\dagger}_i \rho \overline U_i,
\end{eqnarray}
where the $U_i$ are $N$-by-$N$ unitary matrices, chosen at
random from the Haar measure, and the probabilities
$P_i$ are chosen randomly as described
in the Supplemental Equations.  The $P_i$ are all roughly equal.
We pick
\be
1<<D<<N.
\ee

We show in the Supplemental Equations that
\begin{theorem}
\label{mainthm}
For sufficiently large $D$, for sufficiently large $N$, there is a non-zero probability
that a random choice of $U_i$ from the Haar measure and of $P_i$ (as
described in Supplemental Equations) will
give a channel ${\cal E}$ such that
\begin{eqnarray}
H^{\rm min}({\cal E}\otimes \overline {\cal E})&<&
H^{\rm min}({\cal E})+ H^{\rm min}(\overline {\cal E})
\\ \nonumber
&=& 2 H^{\rm min}({\cal E}).
\end{eqnarray}
The size of $N$ depends on $D$.
\end{theorem}

For any pure state input,
the output entropy of ${\cal E}$ is at most $\ln(D)$
and that of ${\cal E} \otimes \overline {\cal E}$ is
at most $2\ln(D)$.
To show theorem (\ref{mainthm}), we first
exhibit an entangled state with a lower output entropy for the channel
${\cal E} \otimes \overline {\cal E}$.
The entangled state we use is the
maximally entangled state:
\be
|\ME\rangle=(1/\sqrt{N})\sum_{\alpha=1}^N |\alpha\rangle\otimes|\alpha\rangle.
\ee
As shown in Lemma 1 in the Supplemental Equations, the output entropy
for this state is bounded by
\be
\label{noD}
H\Bigl({\cal E}\otimes \overline {\cal E}(|\ME\rangle\langle\ME|)\Bigr)
\leq
2\ln(D)-\ln(D)/D.
\ee
  
We then use the random properties of the channel
to show that no product state input can obtain such a low output entropy.
Lemmas 2-5 in the
Supplemental Equations show that, with non-zero probability, the
entropy
$H^{\rm min}({\cal E})$ is at least 
$\ln(D)-\dSm$, for
\be
\label{dsm}
\dSm=c_1/D
+p_1(D){\cal O}(\sqrt{\ln(N)/N})),
\ee
where $c_1$ is a constant and $p_1(D)={\rm poly}(D)$.
Thus, since for large enough $D$, for large enough $N$ we have
$2\dSm< \ln(D)/D$, the theorem follows.

The output entropy can be understood differently:
for a given pure state input, can
we determine from the output which of the unitaries $U_i^{\dagger}$ was applied?
Recall that
\be
\label{ind}
U^{\dagger} \otimes \overline U^{\dagger}|\ME\rangle=|\ME\rangle.
\ee
for any unitary $U$.
This means that,
for the maximally entangled state, if a unitary $U_i^{\dagger}$ was applied to
one subsystem, and $\overline U_i^\dagger$ was applied to the other subsystem,
we cannot determine which unitary $i$ was applied by looking at the output.
This is the key idea behind Eq.~(\ref{noD}).

Note that the minimum output entropy of ${\cal E}$ must be less than
$\ln(D)$
by an amount at least of order $1/D$.
Suppose $U_1$ and $U_2$ are the two unitaries with the
largest $l_i$.
Choose 
a state $|\psi\rangle$ which is an eigenvector of $U_1 U_2^{\dagger}$.
For this state, we cannot distinguish between the states $U_1^{\dagger}|\psi\rangle$
and $U_2|\psi\rangle$, and so
\be
H^{\rm min}({\cal E})\leq 
\ln(D)-(2/D) \ln(2).
\ee
Our randomized analysis bounds how much further the output
entropy of the channel ${\cal E}$ can be lowered for a random choice of
$U_i$.

Our work raises the question of how strong a violation of
additivity is possible.  The relative violation we have found is numerically
small, but it may be possible to increase this, and to find new situations
in which entangled inputs can be used to increase channel capacity, or
novel situations in which entanglement can be used to protect against
decoherence in practical devices.
The map ${\cal E}$ is similar to that used\cite{rugqe} to construct random
quantum expanders\cite{qe1,qe2}, raising the possibility that deterministic
expander constructions can provide stronger violations of
additivity.

While we have used two different channels, it is also possible to find a single
channel ${\cal E}$ such that $H^{\rm min}({\cal E} \otimes {\cal E})<2 H^{\rm min}({\cal E})$,
by choosing $U_i$ from the orthogonal group.
Alternately, we can add an extra classical input used to ``switch" between
${\cal E}$ and $\overline {\cal E}$, as suggested to us by P. Hayden.

The equivalence of the different additivity conjecture\cite{equiv} means that
the violation of any one of the conjectures has profound impacts.  The violation
of additivity of the Holevo capacity means that the problem of
channel capacity remains open,
since if a channel is used many times, we must do an intractable optimization
over all entangled inputs to find the maximum capacity.  
However, we conjecture that additivity
holds for all channels of the form
\be
{\cal E}={\cal F} \otimes \overline {\cal F}.
\ee
Our intuition for this conjecture is that we believe that
{\it multi-party} entanglement (between the inputs to
three or more channels) is not useful, because it is
very unlikely for all channels to apply the same unitary;
note that the state $\ME$
has a low minimum output entropy precisely because it is left
unchanged as in Eq.~(\ref{ind}) if both channels apply corresponding unitaries.
This {\it two-letter} additivity conjecture would allow us to
restrict our attention to
considering input states
with a bipartite entanglement structure, possibly opening
the way to computing the capacity for arbitrary channels.

{\it Acknowledgments---} 
I thank J. Yard, P. Hayden, and A. Harrow.
This work was supported by U. S. DOE Contract No. DE-AC52-06NA25396.

\begin{figure} {
\includegraphics[width=2.5
in]{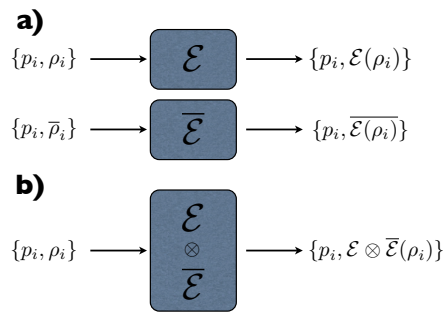} \caption{
Communicating classical information over a quantum channel.
A set of states $\rho_i$ are used with probabilities $p_i$ as signal
states on the channel.  In (a), we use input states which are unentangled
between channels ${\cal E}$ and $\overline {\cal E}$.  In (b), we allow
entanglement.  The capacity of ${\cal E}$ is equal to $\overline {\cal E}$.
The question addressed is whether entangling, as shown in (b), can increase
this capacity. }
} \end{figure}

\begin{figure} {
\includegraphics[width=2.5
in]{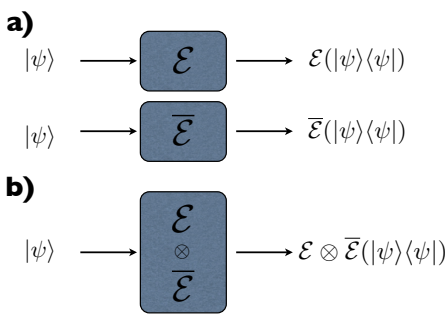} \caption{
Minimum output entropy of a quantum channel.  A pure state
$|\psi\rangle$ is input to the channel.  While the input is a pure state,
the output may be a mixed state.  We attempt to minimize the
entropy of the output state over all pure input states.  The question addressed
is whether an entangled input state, as shown in (b), can have a lower output
entropy for channel ${\cal E} \otimes \overline {\cal E}$, than the sum
of the minimum output entropies for the two channels.  }
} \end{figure}

\end{multicols}

\newpage
{\bf Supplemental Equations:}

To choose the $P_i$, we first choose a set of amplitudes $l_i$ as follows.
For $i=1,...,D$ pick $l_i\geq 0$ independently
from a probability distribution with
\be
\label{rdist}
P(l_i) \propto l_i^{2N-1} \exp(-NDl_i^2),
\ee
where the proportionality constant is chosen such that $\int_0^{\infty} P(l_i)
{\rm d}l_i=1$.  This distribution is the same as that of the length of
a random vector chosen from a Gaussian distribution in $N$ complex dimensions.
Then, define
\be
L=\sqrt{\sum_i l_i^2}.
\ee
Then we set
\be
P_i=l_i^2/L^2,
\ee
so that
\begin{eqnarray}
\label{echannel}
{\cal E}(\rho)&=&
\sum_{i=1}^D 
\frac{l_i^2}{L^2}
U^{\dagger}_i \rho U_i,
\\ \nonumber
\overline {\cal E}(\rho)&=&
\sum_{i=1}^D 
\frac{l_i^2}{L^2}
\overline U^{\dagger}_i \rho \overline U_i,
\end{eqnarray}
The only reason
in what follows for not choosing all the probabilities equal to $1/D$
is that the choice we made will allow us to appeal to certain exact results
on random bipartite states later.

We also define the conjugate channel
\be
\label{ccd}
\conc(\rho)=\sum_{i=1}^D \sum_{j=1}^D
\frac{l_i l_j}{L^2}{\rm Tr}\Bigl(U^{\dagger}_i \rho U_j\Bigr)
|i\rangle\langle j|,
\ee
As shown in \cite{cc}
\be
H^{\rm min}({\cal E})=H^{\rm min}(\conc).
\ee

In the ${\cal O}(...)$ notation that follows, we will take
\be
1<<D<<N.
\ee
We use ``computer science" big-O notation throughout, rather than ``physics"
big-O notation.  That is, if we state that a quantity is ${\cal O}(N)$, it
means that it is asymptotically bounded by a constant times $N$, and
may in fact be much smaller.  For example, $\sqrt{N}$ is ${\cal O}(N)$
in computer science notation but not in physics notation.

Theorem 1
follows from two lemmas below, \ref{lowent} and
\ref{highent}, which give small corrections to the naive estimates
of $2\ln(D)$ and $\ln(D)$
for the entropies.  Lemma \ref{lowent} upper bounds
$H^{\rm min}({\cal E}\otimes \overline {\cal E})$ by
$2\ln(D)-\ln(D)/D$.  Lemma \ref{highent}
shows that for given $D$, for sufficiently large
$N$, with non-zero probability, the
entropy
$H^{\rm min}({\cal E})$ is at least 
$\ln(D)-\dSm$, for
\be
\dSm=c_1/D
+p_1(D){\cal O}(\sqrt{\ln(N)/N})),
\ee
where $c_1$ is a constant and $p_1(D)={\rm poly}(D)$.
Thus, since for large enough $D$, for large enough $N$ we have
$2\dSm< \ln(D)/D$, the theorem follows.

\begin{lemma}
\label{lowent}
For any $D$ and $N$, we have
\begin{eqnarray}
\label{cres}
H^{\rm min}({\cal E}\otimes \overline {\cal E})&\leq &
\frac{1}{D} \ln(D)+\frac{D-1}{D} \ln(D^2)
\\ \nonumber
&=&
2\ln(D)-\frac{1}{D} \ln(D).
\end{eqnarray}
\begin{proof}
Consider the maximally entangled state, $|\ME\rangle=(1/\sqrt{N}) \sum_{\alpha=1}^N
|\alpha\rangle\otimes |\alpha\rangle$.  Then,
\begin{eqnarray}
\label{ec}
{\cal E}\otimes\overline{\cal E}(|\ME\rangle\langle\ME|)&=&
\Bigl(\sum_i \frac{l_i^4}{L^4}\Bigr)
|\ME\rangle\langle\ME| \\ \nonumber
&& +\sum_{i \neq j}
\Bigl(\frac{l_i^2 l_j^2}{L^4}\Bigr)
\Bigl( U^{\dagger}_i \otimes \overline U^{\dagger}_j\Bigr)
|\ME\rangle\langle\ME|
\Bigl( U_i \otimes \overline U_j\Bigr).
\end{eqnarray}
Since the states $|\ME\rangle\langle\ME|$ and
$(U^{\dagger}_i \otimes \overline U^{\dagger}_j)
|\ME\rangle\langle\ME|
(U_i \otimes \overline U_j)$ are pure states, the entropy of
the state in (\ref{ec}) is bounded by
\be
\label{entphibnd}
H({\cal E}\otimes\overline{\cal E}(|\ME\rangle\langle\ME|))
\leq
-\Bigl(\sum_i \frac{l_i^4}{L^4}\Bigr) \ln(
\sum_i \frac{l_i^4}{L^4})
-\sum_{i\neq j}
\Bigl(\frac{l_i^2 l_j^2}{L^4}\Bigr)
\ln(\frac{l_i^2 l_j^2}{L^4})
.
\ee
To show Eq.~(\ref{entphibnd}),
let $\rho_{ij}=
 U^{\dagger}_i \otimes \overline U^{\dagger}_j
|\ME\rangle\langle\ME|
 U_i \otimes \overline U_j$.  Note that
$\rho_{ii}=\rho_{jj}=|\ME\rangle\langle\ME |$
for all $i,j$.  Then, the entropy is equal to
\begin{eqnarray}
\label{entphibnd2}
H({\cal E}\otimes\overline{\cal E}(|\ME\rangle\langle\ME|))
&=&
-{\rm Tr}\Bigl[\Bigl(\sum_i \frac{l_i^4}{L^4}\Bigr) |\ME\rangle\langle\ME | \ln\Bigl(
{\cal E}\otimes\overline{\cal E}(|\ME\rangle\langle\ME|)
\Bigr)\Bigr]
\\ \nonumber
&&-
\sum_{i\neq j}{\rm Tr}\Bigl[\frac{l_i^2 l_j^2}{L^4} \rho_{ij} \ln\Bigl(
{\cal E}\otimes\overline{\cal E}(|\ME\rangle\langle\ME|)\Bigr)\Bigr].
\end{eqnarray}
Using the fact that the logarithm is an operator monotone function\cite{opmon},
we find
that $\ln\Bigl(
{\cal E}\otimes\overline{\cal E}(|\ME\rangle\langle\ME|)\Bigr)
\geq \ln(
\sum_i \frac{l_i^4}{L^4}|\ME\rangle\langle\ME|)$,
and also
that $\ln\Bigl(
{\cal E}\otimes\overline{\cal E}(|\ME\rangle\langle\ME|)\Bigr)
\geq \ln(
\frac{l_i^2 l_j^2}{L^4} \rho_{ij})$ for all $i,j$.
Inserting these inequalities into Eq.~(\ref{entphibnd2}), we
arrive at Eq.~(\ref{entphibnd}).

We claim that the right-hand side of Eq.~(\ref{entphibnd}) is bounded by
\be
\label{hold}
-\Bigl(\sum_i \frac{l_i^4}{L^4}\Bigr) \ln(
\sum_i \frac{l_i^4}{L^4})
-\sum_{i\neq j}
\Bigl(\frac{l_i^2 l_j^2}{L^4}\Bigr)
\ln(\frac{l_i^2 l_j^2}{L^4})
\leq
\frac{1}{D} \ln(D)+\frac{D-1}{D} \ln(D^2).
\ee
To show Eq.~(\ref{hold}), 
define 
$\Pc=\sum_i l_i^4/L^4$.  We claim that
$\Pc\geq 1/D$.  To see this, consider
the real vectors $(l_1^2/L^2,...,l_D^2/L^2)$ and
$(1,...,1)$.  The inner product of these vectors is
equal to $1$ since $\sum_i l_i^2/L^2=1$ while the norms
of the vectors are $\sqrt{\Pc}$ and $\sqrt{D}$, respectively.  Applying the
Cauchy-Schwarz inequality to this inner product, we find
that $\Pc\geq 1/D$ as claimed.
Then the left-hand side of Eq.~(\ref{hold}) is equal to
\begin{eqnarray}
\label{ism}
-\Bigl(\sum_i \frac{l_i^4}{L^4}\Bigr) \ln(
\sum_i \frac{l_i^4}{L^4})
-\sum_{i\neq j}
\Bigl(\frac{l_i^2 l_j^2}{L^4}\Bigr)
\ln(\frac{l_i^2 l_j^2}{L^4})
&=&
-\Pc\ln(\Pc)-(1-\Pc)\ln(1-\Pc)
\\ \nonumber && -(1-\Pc)\sum_{i\neq j}
\Bigl(\frac{l_i^2 l_j^2}{(1-\Pc)L^4}\Bigr)\ln
\Bigl(\frac{l_i^2 l_j^2}{(1-\Pc)L^4}\Bigr)
\\ \nonumber
&\leq &
-\Pc\ln(\Pc)-(1-\Pc)\ln(1-\Pc)
\\ \nonumber && +(1-\Pc)\ln(D^2-D).
\end{eqnarray}
The last line of
Eq.~(\ref{ism}) is maximized at $\Pc=1/D$, giving Eq.~(\ref{hold}),
which implies Eq.~(\ref{cres}).
\end{proof}
\end{lemma}

\begin{lemma}
Consider a random bipartite pure state $|\psi\rangle\langle\psi|$
on a bipartite system with subsystems
$B$ and $E$ with dimensions $N$ and $D$ respectively.
Let $\rho_E$ be the reduced density matrix on $E$.  Then,
the probability density
that $\rho_E$ has a given set of eigenvalues, $p_1,...,p_D$,
is bounded by
\begin{eqnarray}
\label{small2}
&&\tilde P(p_1,...,p_D)\prod_i {\rm d}p_i
\\ \nonumber
&=&
{\cal O}(N)^{{\cal O}(D^2)} 
D^{(N-D)D} \delta(1-\sum_{i=1}^D p_i) \prod_{i=1}^D p_i^{N-D} {\rm d}p_i.
\\ \nonumber
&=& {\cal O}(N)^{{\cal O}(D^2)} 
\delta(1-\sum_{i=1}^D p_i) \prod_{i=1}^D \FF(p_i) {\rm d}p_i,
\end{eqnarray}
where we define
\be
\FF(p)=D^{N-D} p^{N-D} \exp[-(N-D)D(p-1/D)] 
\ee
Note that
$\FF(p)\leq 1$
for all $0\leq p \leq 1$.

Similarly, consider a random state pure state $\rho=|\chi\rangle\langle\chi|$ on
an $N$ dimensional space, and a channel $\conc(...)$ as defined in Eq.~(\ref{echannel}), with
unitaries $U_i$ chosen randomly from the Haar measure and the numbers $l_i$ chosen as
described in Eq.~(\ref{rdist}) and with $N>D$.  Then, the probability density that the eigenvalues of
$\conc(\rho)$ assume given values $p_1,...,p_D$ is bounded by
the same function
$\tilde P(p_1,...,p_D)\prod_i {\rm d}p_i$ as above.

\begin{proof}
As shown in\cite{pl,pl2}, the exact probability distribution of eigenvalues
is
\be
P(p_1,...,p_D)\prod_i {\rm d}p_i\propto 
\delta(1-\sum_{i=1}^D p_i) \prod_{1\leq j<k\leq D} (p_j-p_k)^2
\prod_{i=1}^D p_i^{N-D} {\rm d}p_i,
\ee
where the constant of proportionality is given by the requirement that the probability
distribution integrate to unity.  The proportionality constant is ${\cal O}(N)^{{\cal O}(D^2)} D^{(N-D)D}$ as we show below,
and for $0\leq p_i \leq 1$
\be
\prod_{1\leq j<k\leq D} (p_j-p_k)^2
\prod_{i=1}^D p_i^{N-D}\leq
\prod_{i=1}^D p_i^{N-D},
\ee
so Eq.~(\ref{small2}) follows.
The second equality in (\ref{small2}) holds because $\sum_i (p_i-1/D)=0$.

Given a random pure state $|\chi\rangle\langle\chi|$, with $U_i$ and $l_i$ chosen as
described above, then the state $\conc(|\chi\rangle\langle\chi|)$ has the same eigenvalue distribution
as the reduced density matrix of a random bipartite state, so the second
result follows.  To see that the eigenvalue distribution of
a random bipartite state in $DN$ dimensions is indeed the same as that of
$\conc(|\chi\rangle\langle\chi|)$, we consider the reduced density
matrix on the $N$ dimensional system of the random bipartite state and
show that it has the same statistical properties as
${\cal E}(|\chi\rangle\langle\chi|)$.  We choose the $DN$ different
amplitudes of the 
unnormalized bipartite
state from a Gaussian distribution.
Equivalently, for each $i=1,...,D$ corresponding to a given
state in the environment, we choose an $N$ dimensional
vector $|v_i\rangle$
from a Gaussian distribution.  Thus, before normalization,
the reduced density matrix of the random bipartite state on the $N$
dimensional system has the same statistics as the sum $\sum_{i=1}^D
|v_i\rangle\langle v_i|$ where the $|v_i\rangle$ are states
drawn from a Gaussian distribution.  The state
${\cal E}(|\chi\rangle\langle\chi|)$
is the sum $\sum_{i=1}^D (l_i^2/L^2) U_i^{\dagger} |\chi\rangle\langle\chi|
U_i$.  The
$l_i^2$ have the same statistics as $|v_i|^2$, while the
directions of the vectors $U_i^{\dagger}|\chi\rangle$ are independent
and uniformly distributed, as are the directions of the $|v_i\rangle$.
The factor of $L^2$
takes into account the normalization, so that
${\cal E}(|\chi\rangle\langle\chi|)$ indeed has the same statistics
as the normalized bipartite state as claimed.

Finally, we show how to upper bound the proportionality constant.  One
approach is to keep track of constant factors of $N$ in the derivation of
\cite{pl,pl2}.  Another approach, which we explain here, is to lower bound
the integral
$\int 
\delta(1-\sum_{i=1}^D p_i) \prod_{1\leq j<k\leq D} (p_j-p_k)^2
\prod_{i=1}^D p_i^{N-D} {\rm d}p_i$.  As a lower bound on the integral,
we restrict to a subregion of the integration domain: we assume that
the $i$-th eigenvalue $p_i$ falls into a narrow interval of width $1/N$,
and we choose these intervals such that $|p_i-p_j|\geq 1/N$ for $i \neq j$
and such that
$|p_i-1/D|\leq {\cal O}(D)/N$.  To do this, for example, we can require that the
$i$-th eigenvalue $p_i$ obey $1/D+(2i-D-3/2)/N\leq p_i \leq 1/D+(2i-D-1/2)/N$.
Then, in this subregion, $\prod_{1\leq j<k\leq D} (p_j-p_k)^2\geq
(1/N)^{D^2}$, and $\prod_{i=1}^D p_i^{N-D} \geq (1/D-{\cal O}(D/N))^{(N-D)D}$.
The centers of the intervals were chosen such that if each eigenvalue is
at the center, then $\sum_i p_i=1$; we can then estimate the volume of
the subregion as $\approx \sqrt{1/D} (1/N)^{D-1}$.  Combining these
estimates, we lower bound the integral as desired.
\end{proof}
\end{lemma}

{\bf Remark}:
In order to get some understanding of the probability of having
a given fluctuation in the entropy, we consider a Taylor expansion
about $p_i=1/D$.  The next three paragraphs are not intended
to be rigourous and are not used in the later proof.
Instead, they are intended to, first, give some
rough idea of the probability of a given fluctuation in the entropy,
and, second, explain why $\epsilon$-nets do not suffice to give sufficiently
tight bounds on the probability of having a given fluctuation in the
entropy and hence why we turn to a slightly more complicated way of
estimating this probability in lemmas 3-5.

If all the probabilities $p_i$ are close to $1/D$, so that $p_i=1/D+\delta p_i$ for
small $\delta p_i$, we can 
Taylor expand the last
line of (\ref{small2}),
using $p_i^{N-D}=\exp[(N-D)\ln(p_i)]$, to get:
\be
\label{tpA}
\tilde P(p_1,...,p_D)\approx
{\cal O}(N)^{{\cal O}(D^2)} \exp[-(N-D) D^2 \sum_i \delta p_i^2/2+...].
\ee
Similarly, we can expand
\be
\label{sA}
S=-\sum_i p_i \ln(p_i) \approx \ln(D)-D \sum_i \delta p_i^2/2+...
\ee
Using Eq.~(\ref{tpA},\ref{sA}), we find that the probability of having
$S=\ln(D)-\delta S$ is roughly 
${\cal O}(N)^{{\cal O}(D^2)} \exp[-(N-D) D \delta S]$.

Using $\epsilon$-nets, these estimates (\ref{tpA},\ref{sA}) give
some motivation for the construction we are using, but just fail to
give a good enough bound on their own: define an $\epsilon$-net with
distance $d<<1$ between points on the net.  There are then ${\cal O}(d^{-2N})$
points in the net.  Then, the probability
that, for a random $U_i,l_i$, 
at least one point on the net has a given $\delta S$ is bounded by
$\approx\exp[-N D \delta S+2N \ln(1/d)]$.  Thus, the probability
of having a $\delta S=\ln(D)/2D$ is less than one for
$d\geq D^{-1/4}$.  However, in order to use $\epsilon$-nets to show
that it is unlikely to have any state $|\psi\rangle$ with given
$\delta S$, we need to take a sufficiently dense $\epsilon$-net.
If there exists a state $|\psi^0\rangle$ with given
$\delta S^0$, then any state within distance $d$ will have,
by Fannes inequality\cite{fannes}, a $\delta S\geq \delta S^0-d^2 \ln(D/d^2)$,
and therefore we will need to take a $d$ of roughly $1/\sqrt{D}$
in order to use the
bounds on $\delta S$ for points on the net to
get bounds on $\delta S^0$ with an accuracy
${\cal O}(1/D)$.

However,
in fact this Fannes inequality estimate is usually an overestimate of
the change in entropy.  Given a state $|\psi^0\rangle$ with
a large $\delta S^0$, random nearby states $\chi$ can be written as
a linear combination of $|\psi^0\rangle$ with a random
orthogonal vector $|\phi\rangle$.  Since $\conc(|\phi\rangle\langle\phi|)$ will
typically by close to a maximally mixed state for
random $|\phi\rangle$, and
typically will also have almost vanishing trace with $\conc(|\psi^0\rangle\langle\psi^0|)$,
the state $\conc(|\chi\rangle\langle\chi|)$ will typically be close to a mixture
of 
$\conc(|\psi^0\rangle\langle\psi^0|)$ with the maximally mixed state, and
hence will also have a relatively large $\delta S$.  This idea motivates
what follows.

{\bf Definitions:}
We will
 say that a density matrix $\rho$ is ``close to maximally mixed"
if the eigenvalues 
$p_i$ of $\rho$ all
obey 
\be
|p_i-1/D|\leq \cmm \sqrt{\ln(N)/(N-D)},
\ee
where the constant $\cmm$ will be chosen later.
For any given channel $\conc$, let $P_{\conc}$ denote the probability that,
for a randomly chosen $|\chi\rangle$,
the density matrix $\conc(|\chi\rangle\langle\chi|)$ is close to maximally mixed.
Let $Q$ denote the probability that a random choice of $U_i$ from the Haar measure and
a random choice of numbers $l_i$ produces a channel $\conc$
such that $P_{\conc}$ is less than $1/2$.  Note: we are defining $Q$ to be the
probability of a probability here.
Then,
\begin{lemma}
For an appropriate choice of $\cmm$, the probability $Q$ can be
made arbitrarily close to zero
for all sufficiently large $D$ and $N/D$.
\begin{proof}
The probability $Q$ is less than or equal to $2$ times the probability that for
a random $U_i$, random $l_i$, and random $|\chi\rangle$, the
density matrix ${\conc}(|\chi\rangle\langle\chi|)$ is not close to maximally mixed.
From (\ref{small2}), and as we will explain further in the next
paragraph, this probability is bounded by the maximum over
$p$ such that $|p-1/D|>\cmm \sqrt{\ln(N)/(N-D)}$ of
\begin{eqnarray}
\label{taware}
&&{\cal O}(N^2)^{{\cal O}(D^2)} \FF(p) \\ \nonumber
&=&{\cal O}(N^2)^{{\cal O}(D^2)}
D^{N-D} p^{N-D} \exp[-(N-D)D(p-1/D)] \\ \nonumber
&\approx &\exp[{\cal O}(D^2) \ln(N)-(N-D)D^2
\cmm^2 (\ln(N)/(N-D))/2+...].
\end{eqnarray}
By picking $\cmm$ large
enough, we can make this probability
\be
\label{aware}
{\rm max}_{p,|p-1/D|>\cmm \sqrt{\ln(N)/(N-D)}}
\Bigl({\cal O}(N^2)^{{\cal O}(D^2)} \FF(p)\Bigr)
\ee
 arbitrarily small for sufficiently
large $D$ and $N/D$.

The fact that $\FF(p)\leq 1$ for all $0\leq p \leq 1$ is important in the
claim that (\ref{aware}) indeed is a bound on the given probability.
To compute the probability density for a given set of
eigenvalues, $p_i$, such that for some $j$ we have $|p_j-1/D|>\cmm \sqrt{\ln(N)/(N-D)}$,
we can use the bound $\FF(p)\leq 1$ to show that
${\cal O}(N^2)^{{\cal O}(D^2)} \prod_{i=1}^D \FF(p_i) {\rm d}p_i$ is bounded
by 
${\cal O}(N^2)^{{\cal O}(D^2)} \FF(p_j) \prod_{i=1}^D {\rm d}p_i$.
Therefore,
Eq.~(\ref{aware}) gives a bound on the probability {\it density}
under the assumption that for some $j$ we have $|p_j-1/D|>\cmm \sqrt{\ln(N)/(N-D)}$.

To turns this bound on the probability density into a bound on the
probability, note that
the total integration volume $\int \delta(1-\sum_{i=1}^D p_i)
\prod_{i=1}^D {\rm d}p_i$ is bounded by unity, and the set of
$p_i$ such that for some $j$ we have $|p_j-1/D|>\cmm \sqrt{\ln(N)/(N-D)}$ is a subset
of the set of all $p_i$.

Finally, note that the maximum of Eq.~(\ref{aware}) is achieved
at $|p-1/D|=\cmm \sqrt{\ln(N)/(N-D)}$ and it is straightforward to control the higher terms in the
Taylor expansion of (\ref{taware}) in that case.
\end{proof}
\end{lemma}

The next lemma is the crucial step.
\begin{lemma}
\label{cstep}
Consider a given choice of $U_i$ and $l_i$ which give
a $\conc$ such that $P_{\conc}\geq 1/2$.  
Suppose there exists a state
$|\psi^0\rangle$, such that $\conc(|\psi^0\rangle\langle\psi^0|)$ has
given eigenvalues $p_1,...,p_D$.  Let $\pN$ denote
the probability that, for a randomly
chosen state $|\chi\rangle$, the density matrix $\conc(|\chi\rangle\langle\chi|)$ has
eigenvalues $q_1,...,q_D$ which obey
\be
\label{from}
|q_i-(y p_i+(1-y)(1/D))|\leq 
{\rm poly}(D)
{\cal O}(\sqrt{\ln(N)/(N-D)})
\ee
for some $y\geq 1/2$.
Then,
\be
\label{claim}
\pN \geq \exp(-{\cal O}(N)) (1/2-1/{\rm poly}(D)),
\ee
where the power of $D$ in the polynomial in (\ref{claim}) can be made arbitrarily large by
an appropriate choice of the polynomial in (\ref{from}).
\begin{proof}
Consider a random state $\chi$.
We can write $|\chi\rangle$ as a linear combination
of $|\psi^0\rangle$ and a state $|\phi\rangle$ which is
orthogonal to $\psi^0\rangle$ as follows:
\be
\label{below}
|\chi\rangle=z \sqrt{1-x^2}|\psi^0\rangle+x|\phi\rangle,
\ee
where $z$ is a phase: $|z|=1$.

For random $\chi$, the probability that $x^2\leq 1/2$ is
$\exp(-{\cal O}(N))$.  We can also calculate this
probability exactly.  Let
$S_n$ be the surface area of a unit hypersphere in $n$ dimensions.
Then, the probability that $x^2\leq 1/2$ is equal to
\begin{eqnarray}
\label{below2}
&&S_{2N}^{-1} \int_0^{\pi/4} 
2\pi \cos(\theta) \sin(\theta)^{2N-3} S_{2N-2} \,
{\rm d\theta} 
\\ \nonumber
&=& 
(1/\sqrt{2})^{2N-2}
\\ \nonumber
&=&\exp[-\ln(2) (N-1)]
\\ \nonumber
&=& \exp(-{\cal O}(N)).
\end{eqnarray}

Since $\chi$ is random,
the probability distribution of $|\phi\rangle$ is that of a random state with
$\langle \phi|\psi^0 \rangle=0$.  One way to generate such a random state $|\phi\rangle$
with this property is to choose a random state $|\theta\rangle$ and set
\be
|\phi\rangle=\frac{1}{\sqrt{1-|\langle\psi^0|\theta\rangle|^2}}
\Bigl( 1-|\psi^0\rangle\langle\psi^0| \Bigr) |\theta\rangle.
\ee
If we choose a random state $|\theta\rangle$,
then with probability at least $1/2$, the state $\conc(|\theta\rangle\langle\theta|)$ is
close to maximally mixed.  Further, for any given $i,j$,
the probability that $|\langle \psi^0|U_i U_j^{\dagger} |\theta\rangle|$ is greater than
${\cal O}(\sqrt{\ln(D)/N})$ is $1/{\rm poly}(D)$, and the polynomial
${\rm poly}(D)$ can be chosen to be any given power of $D$ by appropriate choice of the
constant hidden in the ${\cal O}$ notation for
${\cal O}(\sqrt{\ln(D)/N})$.
Therefore,
\be
\label{smalltrCross}
{\rm Pr}\Bigl[{\rm Tr}\Bigl(|\conc(|\theta\rangle\langle\psi^0|)|\Bigr) \geq {\rm poly}(D)
\sqrt{\ln(D)/N}\Bigr] \leq 1/{\rm poly}(D),
\ee
with any desired power of $D$ in
the polynomial on the right-hand side (the notation ${\rm Tr}(|...|)$ is
used to denote the trace norm here).

Then, since 
\be
\label{closer}
{\rm Pr}\Bigl[\Bigl||\phi\rangle-|\theta\rangle\Bigr|\geq
{\cal O}(\sqrt{\ln(D)/N})\Bigr]\leq 1/{\rm poly}(D),
\ee
we find that
\begin{eqnarray}
\label{noTr}
&& {\rm Pr}\Bigl[{\rm Tr}\Bigl(|\conc(|\phi\rangle\langle\psi^0|)|\Bigr) 
\geq 
 {\rm poly}(D)
\sqrt{\ln(D)/N}\Bigr] 
\nonumber \\
&\leq & 1/{\rm poly}(D),
\end{eqnarray}
with again any desired power in the polynomial.

The probability that 
$\conc(|\theta\rangle\langle\theta|)$ is close to maximally mixed is
at least $1/2$, and so by (\ref{ccd},\ref{closer}) the probability that
the eigenvalues $r_1,...,r_D$
of $\conc(|\phi\rangle\langle\phi|)$ obey
\begin{eqnarray}
|r_i-1/D|& \leq &
\cmm \sqrt{\ln(N)/(N-D)}+{\rm poly}(D)(\ln(D)/N) \nonumber \\
&\leq & {\rm poly}(D) {\cal O}(\sqrt{\ln(N)/N})
\end{eqnarray}
is at least $1/2-1/{\rm poly}(D)$.
Let
\be
y=1-x^2.
\ee
Thus,
since
\begin{eqnarray}
\conc(|\chi\rangle\langle\chi|)&=&(1-x^2) \conc(|\psi^0\rangle\langle\psi^0|) + x^2 
\conc(|\phi\rangle\langle\phi|)\\ \nonumber
&&+ \Bigl(\overline z x\sqrt{1-x^2} \conc(|\phi\rangle\langle\psi^0|)+h.c.\Bigr),
\end{eqnarray}
using Eq.~(\ref{noTr})
we find that for given $x$, 
the probability that
a randomly chosen $|\phi\rangle$ gives a state with
eigenvalues
$q_1,...,q_D$ such that
\be
\label{aeq}
|q_i-(yp_i+(1-y)(1/D))|\leq 
{\rm poly}(D)
{\cal O}(\sqrt{\ln(N)/N})
\ee
is $1/2-1/{\rm poly}(D)$.  Combining this result with the $\exp(-{\cal O}(N))$
probability of $x^2\leq 1/2$, the claim of the lemma follows.
\end{proof}
\end{lemma}

We now give the last lemma which shows a lower bound, with non-zero probability,
on $H^{min}(\conc)$.  The basic idea of the proof is to estimate
the probability that a random state
input into a random channel $\conc$ gives an output state with moderately
low output entropy (defined slightly differently below in terms
of properties of the eigenvalues of the output density matrix).  We estimate
this probability in two different ways.  First, we estimate the probability
of such an output state conditioned on $\conc$ being chosen such
that there exists some input state
with an output entropy less than $\ln(D)-\dSm$.  Next, we
estimate the probability of such an output state, without any conditioning
on $\conc$.
By comparing these estimates, we are able to bound the probability of $\conc$
having an input state which gives an output entropy less than $\ln(D)-\dSm$.
\begin{lemma}
\label{highent}
If the unitary matrices $U_i$ are chosen at random from the Haar measure,
and the $l_i$ are chosen randomly as described above, then
the probability that $H^{min}(\conc)$ is less than
$\ln(D)-\dSm$
is less than one for sufficiently large $N$, for appropriate choice of
$c_1$ and $p_1$.
The $N$ required depends on $D$.
\begin{proof}
Let $\Pb$ denote the probability that $H^{min}(\conc)<\ln(D)-\dSm$.
Then, with probability at least $\Pb-Q$, for
random $U_i$ and $l_i$,
the channel
${\conc}$ has $P_{\conc}\geq 1/2$ and has $H^{min}(\conc)<\ln(D)-\dSm$.

Let $|\psi^0\rangle$ be a state
which minimizes the output entropy of channel $\conc$.
By lemma \ref{cstep}, for such a channel, for a random state
$|\chi\rangle$, the density matrix $\conc(|\chi\rangle\langle\chi|)$ has
eigenvalues $q_1,...,q_D$ which obey
\be
\label{noway}
|q_i-(yp_i+(1-y)(1/D))|\leq 
{\rm poly}(D)
{\cal O}(\sqrt{\ln(N)/N})
\ee
for $y\geq 1/2$ with probability at least
\be
\exp(-{\cal O}(N)) (1/2-1/{\rm poly}(D)).
\ee

Therefore, for a random choice of $U_i,l_i,\chi$, the
state $\conc(|\chi\rangle\langle\chi|)$ has eigenvalues
$q_i$ which obey Eq.~(\ref{noway}) with probability at
least
\be
\label{Pgeq}
(\Pb-Q) \exp(-{\cal O}(N)) (1/2-1/{\rm poly}(D)).
\ee

However, by Eq.~(\ref{small2}), the probability of having such eigenvalues
$q_i$ is bounded by the maximum of the probability
density $\tilde P(q_1,...,q_D)$ over $q_i$ which obey
Eq.~(\ref{noway}).  Given the assumptions that
$-\sum_i p_i \ln(p_i)\leq \ln(D)-\dSm$, $y\geq 1/2$, and the constraint that
$\sum_i p_i=1$, 
the quantity 
$\tilde P(q_1,...,q_D) \leq
{\cal O}(N)^{{\cal O}(D^2)}\exp[-c_2 (N-D)]$,
where $c_2$ can be made arbitrarily large by choosing $c_1$ large (the proof
of this statement is given in the next paragraph).
We pick $c_{MM}$ so that $Q<1$ and then if $P_{bad}=1$, we can pick
$c_1$ and $p_1$ such that for sufficiently large
$N$ this quantity $\tilde P(q_1,...,q_D)$ is less than that
in (\ref{Pgeq}), giving a contradiction.
Comparing to the discussion
below Eq.~(\ref{below}), we see that we need $c_2>\ln(2)$ to
get this contradiction.
Therefore, $P_{bad}<1$.  In fact, since $Q$ can be
made arbitrarily close to zero, $P_{bad}$ can be made arbitrarily close
to zero for sufficiently large $D,N$.

Finally, we briefly show how $c_2$ can be made arbitrarily large by
choosing $c_1$ sufficiently large.  The natural way to do this is by
treating this problem as a constrained maximization problem: maximize
the probability $\tilde P(q_1,...,q_D)$ subject to a contraint
on the entropy of the $p_i$.  This maximization can be done with Lagrange
multipliers, and the final result is obtained after
a direct, but slightly lengthy,
calculation.  We now show a slightly different way to obtain the
same result.  First, we claim that we can
find
constants $x,y$ with $0<x<1<y$, such that the probability 
that an eigenvalue $q_i$ falls outside
the interval $(x/D,y/D)$ is bounded by
${\cal O}(N)^{{\cal O}(D^2)} \exp[-c_2 (N-D)]$ for any desired $c_2$.
To show this claim, we use the fact that
this probability is bounded by
${\cal O}(N)^{{\cal O}(D^2)} {\rm max}(\FF(x/D),\FF(y/D))$.
The function $\FF(x/D)=\exp[-(N-D)(x-1)+(N-D)\ln(x)]=\exp[(N-D)(-x+1+\ln(x))]$.
We choose $x$ sufficiently small that $-x+1+\ln(x)\leq c_2$ and similarly
we choose $y$ sufficiently large that $-y+1+\ln(y)\leq c_2$, and then
we have bounded the probability of any eigenvalue $q_i$ lying outside
this interval.  Thus, we can assume that the eigenvalues lie inside this
interval.

Next, for any set of eigenvalues $\{q_i\}$ which all lie in this interval,
we have
\be
\label{worstcaseeig}
\prod_i \FF(q_i)\leq \exp[-(N-D)D^2 \sum_i (q_i-1/D)^2/2y^2].
\ee
Comparing Eq.~(\ref{worstcaseeig}) to Eq.~(\ref{tpA}), we have
worsened by a constant in the exponent ($1/2y^2$ instead of $1/2$), but
the inequality is now valid for all $q_i$ in the interval $(x/D,y/D)$,
not just as a Taylor expansion.
We now also give a bound
on the entropy.
For any set of eigenvalues of the density matrix $p_i$, we have
\begin{eqnarray}
\label{worstcase}
S(\{p_i\})&=&-\sum_i p_i \ln(p_i)
\\ \nonumber
&\equiv&\ln(D)-\delta S
\\ \nonumber
&\geq &\ln(D)-D\sum_i (p_i-1/D)^2.
\end{eqnarray}
To derive Eq.~(\ref{worstcase}), note that
$\sum_i -p_i \ln(p_i)=\ln(D)+\sum_i [(1/D) \ln(1/D)-p_i\ln(p_i)+(p_i-1/D) (\ln(1/D)+1)]$, because $\sum_i (p_i-1/D)=0$.
Then, $\delta S=
-\sum_i [(1/D) \ln(1/D)-p_i\ln(p_i)+(p_i-1/D) (\ln(1/D)+1)]$.  For $p_i=1/D$,
$-[(1/D) \ln(1/D)-p_i\ln(p_i)+(p_i-1/D) (\ln(1/D)+1)]=0$, while for $p_i=0$, it is
equal to $1/D$.  The function $D (p_i-1/D)^2$ is a quadratic function
chosen to fit these two points ($0$ at $p_i=1/D$ and $1/D$ at $0$), and
both $D(p_i-1/D)^2$ and
$(1/D) \ln(1/D)-p_i\ln(p_i)+(p_i-1/D) (\ln(1/D)+1)$ have vanishing
derivative at $p_i=1/D$; it was to make the derivative vanish that
we subtracted off that linear term.
By checking the sign of the third derivative of $-p_i \ln(p_i)$ one
may verify the inequality (\ref{worstcase}).

Comparing Eq.~(\ref{worstcase}) to Eq.~(\ref{sA}), we have lost the factor
of $1/2$ in (\ref{worstcase}), but the result is now an inequality valid for
all $p_i$, not just a Taylor expansion.
Comparing Eq.~(\ref{worstcaseeig}) and Eq.~(\ref{worstcase}),
and using Eq.~(\ref{noway}), we find
\be
\prod_i \FF(q_i)\leq \exp\Bigl\{-\frac{(N-D) D [\delta S-{\rm poly}(D){\cal O}(\sqrt{\ln(N)/N)}]}{8y^2}\Bigr\},
\ee
and so we can make $c_2$ arbitrarily large by choosing sufficiently large $c_1$.

\end{proof}
\end{lemma}
This completes the proof of the theorem.

\end{document}